\documentclass[12pt,preprint]{emulateapj}

\def\be{\begin{equation}}
\def\ee{\end{equation}}
\def\bea{\begin{eqnarray}}
\def\eea{\end{eqnarray}}

\usepackage{graphicx}

\begin{document}

\title{A supra-massive magnetar central engine for short GRB 130603B}

\author{Yi-Zhong Fan$^{1}$, Yun-Wei Yu$^{2}$, Dong Xu$^{3}$, Zhi-Ping Jin$^{1}$, Xue-Feng Wu$^{4}$, Da-Ming Wei$^{1}$ and Bing Zhang$^{5}$}
\affil{$^1$ {Key Laboratory of dark Matter and Space Astronomy, Purple Mountain Observatory, Chinese Academy of Sciences,
Nanjing, 210008, China.}\\
$^{2}$ {Institute of Astrophysics, Central China Normal University, Wuhan 430079, China.}\\
$^{3}$ {Dark Cosmology Centre, Niels Bohr Institute, University of Copenhagen, Juliane Maries Vej 30, 2100 Copenhagen, Denmark.}\\
$^{4}$ {Chinese Center for Antarctic Astronomy, Purple Mountain Observatory,
Chinese Academy of Sciences, Nanjing 210008, China}\\
$^{5}$ Department of Physics and Astronomy, University of Nevada, Las Vegas, NV 89154, USA}
 \email{yzfan@pmo.ac.cn (YZF) and zhang@physics.unlv.edu (BZ)}

\begin{abstract}
We show that the peculiar early optical and in particular X-ray afterglow emission
of the short duration burst GRB 130603B can be explained by
continuous energy injection into the blastwave from a supra-massive
magnetar central engine. The observed energetics and
temporal/spectral properties of the late infrared bump (i.e., the
``kilonova") are also found consistent with emission from the
ejecta launched during an NS-NS merger and powered by a magnetar
central engine.
The isotropic-equivalent kinetic energies of both the GRB blastwave
and the kilonova are about $E_{\rm k}\sim 10^{51}$ erg, consistent with
being powered by a near-isotropic magnetar wind. However, this relatively
small value demands that
{\it most of the initial rotational energy
of the magnetar $(\sim {\rm a~ few \times 10^{52}~ erg})$ is carried away by gravitational wave radiation.}
Our results suggest that (i) the progenitor of GRB 130603B would be
a NS-NS binary system, whose merger product would be a supra-massive
neutron star that lasted for about $\sim 1000$ seconds;
(ii) the equation-of-state of nuclear matter would be stiff enough to
allow survival of a long-lived supra-massive neutron star, so that
it is promising to detect bright electromagnetic counterparts of
gravitational wave triggers without short GRB associations in the
upcoming Advanced LIGO/Virgo era.
\end{abstract}

\setlength{\parindent}{.25in}

\section{Introduction}
Short-duration, hard-spectrum $\gamma$-ray bursts (short GRBs),
whose durations are typically less than two seconds, have been
widely speculated to be powered by mergers of two compact objects,
either two neutron stars, NS-NS; or a neutron star and a black hole,
NS-BH \citep{Eichler1989,Narayan1992}. Tentative evidence for such
progenitor models includes the host galaxy properties, locations in
the host galaxies, as well as the non-association of bright
supernovae with short GRBs \citep{Gehrels2005,Leibler2010}. A
``smoking-gun'' signature of these events are the so-called
kilonova, which is a supernova-like near-infrared/optical transient
powered by radioactive decay of heavy elements synthesized in the
ejecta launched during the mergers
\citep{Li1998,Kulkarni2005,Rosswog2005,Metzger2010} and sometimes
also contributed by a long-lived central engine
\citep{Kulkarni2005,Yu2013}. Such a signal had remained elusive due
to its faint and transient nature until recently.  At 15:49:14 UT on
June 3 2013, GRB 130603B with a duration $T_{90} = 0.18\pm 0.02$ s
in the 15-350 keV band triggered the Burst Alert Telescope (BAT)
onboard the {\it Swift} satellite. This burst is an archetypal
short-hard GRB \citep{deUgartePostigo2013} because of the following
properties: (a) The BAT light curve did not show any ``extended
emission'' down to $\sim 0.005 ~{\rm count~ det^{-1}~s^{-1}}$ level;
(b) a spectral lag analysis revealed no significant lag between low
and high energy photons; (c) Observations of the event by Konus/WIND
gave a rest frame peak energy $E_{\rm peak,rest}=895\pm135$ keV. A
{\em Hubble Space Telescope} (HST) observation was made about 1 week
after trigger, It revealed a bright near-infrared source
\citep{Tanvir2013,Berger2013}, which is suggested to be consistent
with the prediction of the kilonova calculations \citep{Kasen2013}.
This supports the compact star merger origin of this short GRB. Aother
possibility proposed by Jin et al. (2013) that the infrared bump may
be attributed to the synchrotron radiation of a mildly-relativistic
blast wave is disfavored by the non-detection of a simultaneous
brightening in the radio band (Fong et al. 2013).

In this {\it Letter}, we constrain the central engine properties using
both the early X-ray and U-band afterglow and the late infrared bump data.
We argue that the data are consistent with a supra-massive magnetar
that undergoes significant gravitational wave energy loss during
the early spin-down phase.

\section{The early X-ray and U-band afterglow and the late infrared bump: shedding light on the central engine}
\begin{figure}
\begin{picture}(0,300)
\put(0,0){\includegraphics{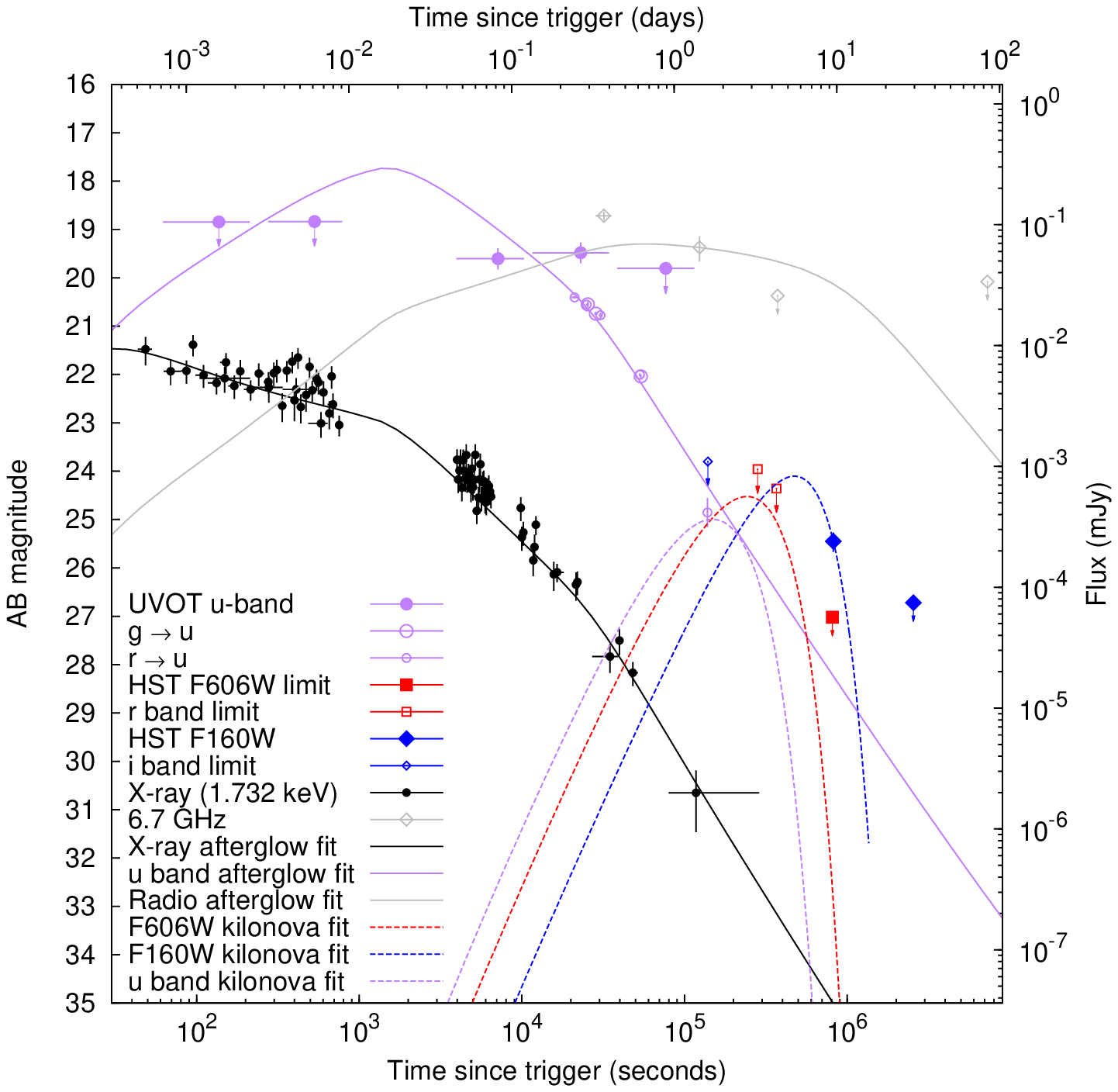}}
\end{picture}
\caption{The broad-band (black: X-ray; purple: u-band, red: r-band, blue: HST F160W; grey: 6.7 GHz) light curves of GRB 130603B and the theoretical model curves. The X-ray data are from http://www.swift.ac.uk/xrt$_{-}$curves/00557310/ \citep{Evans2009}, while the u-band lightcurve is based on the data reported in \citet{deUgartePostigo2013} and \citet{Tanvir2013}, with the g-band and r-band data extrapolated to the UVOT u-band. Proper corrections for extinction in the Milky Way Galaxy and the GRB host galaxy have been made. The solid curves are the theoretical afterglow prediction with a magnetar energy injection. The dotted curves are the kilonova predictions in the u, F606W, and F160W bands, respectively.} \label{fig:1}
\end{figure}

The X-ray telescope (XRT) on board {\em Swift} began to take data from 43 s after the burst trigger. The most remarkable feature of the X-ray data is a shallow decline phase lasting $\sim 1000$ s (Fig.1, adapted from http://www.swift.ac.uk/xrt$_{-}$curves/00557310/ \citep{Evans2009}). {\em Swift}'s Ultraviolet Optical Telescope (UVOT) started to observe the burst approximately 62 s after trigger. A faint source at the location of the afterglow was detected in all 7 UVOT filters but these emissions are so faint that a single U filter light curve was computed from the 7 UVOT filters to improve the signal to noise \citep{deUgartePostigo2013}. This is also presented in Fig.1 (data from \citet{deUgartePostigo2013} except that  the upper limits are at a $3\sigma$ confidence level). The early upper limits and later detections in the U-band suggests that the early U band light curve should be flat or even rise with time. These features are difficult to interpret within the standard fireball afterglow model framework if the central engine energy injection is impulsive. This can be understood as the following. If the GRB central engine gives an impulsive energy injection, a shallowly decaying
X-ray lightcurve and a slowly rising optical lightcurve at early epochs are possible
only if the fireball is in the fast cooling regime \citep{Sari1998}, and the X-ray
band is between the cooling frequency ($\nu_{\rm c}$) and the typical synchrotron
frequency ($\nu_{\rm m}$), while the U-band is below $\nu_{\rm c}$. Such a possibility,
however, is strongly disfavored by the spectral data. The spectrum of X-ray emission
at $t<1000$ s is $F_\nu \propto \nu^{-0.97\pm 0.22}$, which is inconsistent with a
fast cooling spectrum $F_\nu \propto \nu^{-1/2}$. An even much stronger constraint
comes from the spectral energy distribution at $t\sim 0.35$ day, which gives
a $\nu_{\rm c} \approx 10^{16}$ Hz \citep{deUgartePostigo2013,Jin2013}. Together with the 6.7 GHz flux $\sim 0.13$ mJy at
$t\sim 0.37$ day, one has $\nu_{\rm m} \approx 2\times 10^{12}$ Hz and the maximum specific flux $F_{\nu_{\rm max}} \approx 0.8$ mJy (see also Fong et al. 2013). For a burst born in an ISM-like circum-burst medium, we have $\nu_{\rm c}\propto
t^{-1/2}$ and $\nu_{\rm m}\propto t^{-3/2}$ \citep{Sari1998}. Therefore at $t\sim 0.01$
day we have $\nu_{\rm c} \sim 6\times 10^{16}$ Hz and $\nu_{\rm m} \sim 6\times 10^{14}$
Hz, which are far from what is needed in the fast cooling model.

An X-ray shallow decline phase is commonly observed in {\it Swift} long GRBs, which can be interpreted as energy injection into the GRB blast wave from a long-lasting central engine \citep{Zhang2006}.
We take a general energy injection law $dE_{\rm inj}/dt \propto {t}^{-q}$ for $t<t_{\rm end} \sim 10^{3}$ s \citep{Zhang2006,Dai1998,Zhang2001}, which gives $\nu_{\rm m}\propto t^{-(2+q)/2}$, $\nu_{\rm c} \propto t^{(q-2)/2}$ and $F_{\nu, \rm max} \propto t^{1-q}$. Since at $t\sim t_{\rm end}$, one has $\nu_{\rm m}\sim 10^{15}$ Hz and $\nu_{\rm c} \sim 6\times 10^{16}$ Hz, the optical emission should rise with time as $F_{\nu_{\rm opt}} \propto t^{(8-5q)/6}$ while the X-ray (1.7 keV) emission drops with time as $F_{\nu_{\rm X}}\propto t^{[2(2-p)+(p+2)q]/4}$. For $q \sim 0$ relevant for a spinning down magnetar due to magnetic dipole radiation, both the peculiar X-ray and optical emissions can be accounted for (Fig.1). Such a relatively long, steady energy injection with a roughly constant luminosity is difficult to fulfill in the NS-NS or NS-BH merger scenarios with a black hole central engine. Simulations suggest that long-lasting emission may arise in these systems, but the accretion rate history is essentially defined by the accretion rate of fall-back materials, which typically satisfies
$dE_{\rm inj}/dt\propto t^{-5/3}$ \citep[see Fig.3 of][for both NS-NS merger and NS-BH merger scenarios]{Rosswog2007} and is {\it far from what needed in current afterglow modeling\footnote{If the magnitude of the viscosity decreased dramatically from $\alpha \sim 0.1$ to $\sim 10^{-4}$ for the prompt accretion and the fall-back accretion phases, the prompt emission lasting $\sim 0.1$ s and a late plateau-like emission lasting $\sim 10^{3}$ s may be possible (Lee et al. 2009). It is however unclear how such a huge change of $\alpha$ can take place.}}.

{ We then turn to the possibility that a long-lived supra-massive magnetar rather than a black hole was promptly formed after the merger \citep[][]{Gao2006,FanXu2006,Metzger2008,Bucciantini2012,Rowlinson2013,Zhang2013}. This is possible if the neutron star equation-of-state is stiff enough, and if the total mass of the two merging neutron stars is not large enough (e.g., Morrison et al. 2004; Giacomazzo \& Perna 2013).}
Indeed, for a sufficiently stiff equation of state yielding $M_{\rm max}\sim 2.2-2.3~M_\odot$, the merger of double neutron stars with a total gravitational mass $M_{\rm tot}\sim 2.6~M_\odot$ (note that among the ten neutron star binary systems identified so far, five have such a total mass) can produce a supra-massive magnetar with  $P_0 \sim 1$ ms, which survives until a good fraction of its rotational energy has been lost via dipole radiation and gravitational wave radiation \citep[see][and the references therein]{Fan2013}.

{ As already mentioned, at $t\sim 0.35$ day, the key parameters governing the synchrotron spectrum are $\nu_{\rm c}\approx 10^{16}$ Hz, $\nu_{\rm m}\approx 2\times 10^{12}$ Hz and $F_{\nu_{\rm max}}\approx 0.8~{\rm mJy}$. Adopting eqs.(2-4) of Fan \& Piran (2006; see also Yost et al. 2003), it is straightforward to show
\[
E_{\rm k}\approx 1.5\times 10^{51}~{\rm erg}~n_{-1}^{-1/5},~\epsilon_{\rm e}\approx 0.2n_{-1}^{1/5},~\epsilon_{\rm B} \approx 0.04n_{-1}^{-3/5},
\]
where $E_{\rm k}$ is the isotropic-equivalent kinetic energy of the ejecta and $\epsilon_{\rm e}$ ($\epsilon_{\rm B}$) is the fraction of shock energy given to the electrons (magnetic field), $n$ is the number density of the interstellar medium and has been normalized to $0.1~{\rm cm^{-3}}$, and the energy distribution power-law index of the shock electrons is taken as $p= 2.3$ based on the optical and X-ray spectral data \citep[][]{deUgartePostigo2013,Jin2013}. Please note that we have inserted the Compton parameter $Y \approx [-1+\sqrt{1+4(\nu_{\rm m}/\nu_{\rm c})^{(p-2)/2}\epsilon_{\rm e}/\epsilon_{\rm B}}]/2 \sim 1$ into eq.(4) of Fan \& Piran (2006) to estimate the physical parameters governing $\nu_{\rm c}$.

Interestingly these relations imposes a tight constraint on the isotropic-equivalent kinetic energy of the ejecta, i.e., $E_{\rm k}\approx 1.5\times 10^{51}~{\rm erg}~(\epsilon_{\rm B}/0.04)^{1/3}$. It is well known that $\epsilon_{\rm B}$ is not expected to be considerably larger than $\sim 1/3$ (i.e., the equilibrium argument, for which the shock energy is equally shared among electrons, protons, and magnetic fields), we then have \[E_{\rm k}<3\times 10^{51}~{\rm erg}.\] Our result is remarkably consistent with the independent modeling of the late X-ray data by \citet{Fong2013}, in which $E_{\rm k}<1.7\times 10^{51}$ erg was inferred.

Therefore we have shown analytically that the peculiar X-ray and optical data in the first $\sim 1000$ s strongly suggests the energy injection of the magnetar into the blast wave. On the other hand, the normally declining late radio, optical and X-ray afterglow data impose a very tight constraint on the kinetic energy of the blast wave $E_{\rm k} \sim 10^{51}$ erg. The energy injection rate is thus needed to be
\[
dE_{\rm inj}/dt' \sim (1+z)E_{\rm k}/t_{\rm end} \sim 10^{48}~{\rm erg~s^{-1}}.
\]

It is intriguing to note that modeling of the infrared bump also gives a similar isotropic energy for the kilonova component. According to \citet{Tanvir2013} and \citet{Berger2013} and also our numerical results below, the data require  $V_{\rm kilonova}\sim 0.1-0.3c$ and $M_{\rm kilonova}\sim 0.03-0.08M_{\odot}$, which correspond to $E_{\rm kilonova}\sim 10^{51}$ erg. The near constant isotropic equivalent kinetic energy in the jet component and the kilonova component would demand coincidence if the central engine is a black hole, but would be a natural outcome if the central engine is a millisecond magnetar. { We postulate that the magnetar collapses into a black hole at $t_{_{\rm end}} \sim 1000$ s. Interestingly, Fong et al. (2013) suggested that the X-ray afterglow has an unexpected excess at $t>1$ day, which may be attributed to the emission powered by fall-back accretion onto a central black hole. Such a model is in agreement with our scenario.}

\section{Possible gravitational wave losses}

For a long-lived magnetar formed in an NS-NS merger, the initial rotation period is expected to be $P_0\sim 1$ms, and only for such a short spin period the uniform rotation can play an non-ignorable role in stabilizing the magnetar (see Fan et al. 2013 and the references therein). The initial rotational energy of a supremassive magnetar is therefore expected to be a few $\times 10^{52}$ erg, much larger than the inferred $E_{\rm k}\sim 10^{51}$ erg.
Lack of a bright electromagnetic emission component, the missing energy has to be carried away by a non-electromagnetic component. There is no known mechanism to release this huge amount of energy via thermal neutrinos.
Here we focus on the possibility that the energy is carried away via gravitational wave radiation.

A magnetar loses rotational energy through magnetic dipole radiation and gravitational wave radiation \citep{Shapiro1983}
\[
-dE_{\rm rot}/dt'=\pi^{4}R_{\rm s}^6B_{\rm \perp}^{2}f^{4}/6c^{3}+32\pi^{6}GI_{\rm zz}^{2}\epsilon^{2}f^{6}/5c^{5},
\]
where $t'\equiv t/(1+z)$, $\epsilon=2(I_{\rm xx}-I_{\rm yy})/(I_{\rm xx}+I_{\rm yy})$ is the ellipticity in terms of the principal moments of inertia (i.e., $I$), $R_{\rm s}$ is the radius of the magnetar, $B_{\rm s}$ is the surface magnetic field strength {\em at the pole}, and $f=2/P$ ($P$ is the rotation period in units of second). To lose a considerable amount of rotational energy of the supramassive magnetar in $t'_{\rm end}\sim 10^{3}$ s mainly via gravitational wave radiation, the ellipticity should be
\begin{equation}
\epsilon \approx 0.0034~\left({I\over 10^{45.2}{\rm g~cm^{2}}}\right)^{-1/2}\left({P_{0}\over 1{\rm ms}}\right)^{2}\left({t'_{\rm end}\over 10^3~{\rm s}}\right)^{-1/2}.
\label{eq:epsilon_constraint}
\end{equation}
Such an ellipticity is larger than the maximum elastic quadrupole deformation of a conventional neutron star. It may be accommodated by one of the following two possibilities. First, if the magnetar is not a normal neutron star but a crystalline color-superconducting quark matter \citep{Xu2003,Lin2007}, then such a large deformation is allowed ({The quark star model, though highly speculative,} is also helpful to solve the baryon pollution problem of GRBs \citep{Dai1998,Paczynski2005}). Another possibility discussed the literature is that a super-strong interior magnetic field of a magnetar could induce a sizable prolate deformation \citep{DallOsso2009,Mallick2013}. Indeed a recent study on the magnetic field slow decay of neutron stars suggests that the initial interior magnetic field of both Soft Gamma-ray Repeaters and Anomalous X-ray Pulsars is $\gtrsim 10^{16}$ Gauss \citep{DallOsso2012}. In principle, for a magnetar rotating with $P_0\sim 1$ ms, an initial interior magnetic field $\sim 10^{17}$ Gauss is possible \citep{Duncan1992}. The interior field is likely dominated by the toroidal component $B_{\rm t}$ (In a recent calculation, Fujisawa et al. (2012) found that the volume-averaged poloidal component could be $\sim 5B_{\rm s}$). Adopting eq.(4) of Usov (1992)\footnote{If one adopts eq.(8) of \citet{DallOsso2009} the estimated $\epsilon$ would be a few times smaller. However, as summarized in their second footnote, that estimate is a very conservative lower limit.}, we get $\epsilon \sim 0.004$ for $B_{\rm t}\sim 5\times 10^{16}~{\rm G}$. The large required deformation would be possible if the strong toroidal magnetic field is stable \citep[e.g.][]{Braithwaite2008}.}

The dipole radiation of the supramssive magnetar has a luminosity \citep{Shapiro1983}
\begin{equation}
L_{_{\rm dip}}\approx 2.5\times 10^{48}~{\rm erg~s^{-1}}\left({R_{\rm s}\over 10^{6}{\rm cm}}\right)^6 \left({B_{\rm \perp}\over 5\times 10^{14}{\rm G}}\right)^{2} \left({P_0\over 1{\rm ms}}\right)^{-4},
\end{equation}
where $B_{\rm \perp}=B_{\rm s} \sin \alpha$, and $\alpha$ is the angle between the rotational and dipole axes. One can see the required magnetar luminosity ($E_{\rm k}/t'_{\rm end}$) can be reproduced with a reasonable $B_{\perp}\sim 5\times 10^{14}$ Gauss.
For magnetar energy loss dominated by gravitational wave radiation, the energy injection rate into the blast wave and the ejecta launched during the merger can be approximated as $dE_{\rm inj}/dt' \approx L_{\rm dip}(1+t'/\tau'_{_{\rm GW}})^{-1}$, as long as the gravitational wave radiation luminosity is much larger than $L_{\rm dip}$, where $\tau'_{_{\rm GW}} \approx 680~{\rm s}~({I\over 10^{45.2}{\rm g~cm^{2}}})^{-1}({P_{0}\over 1{\rm ms}})^{4} ({\epsilon\over 0.004})^{-2}$ is the spin down timescale of the magnetar due to gravitational wave radiation. For $t'\leq t'_{\rm end} \approx \tau'_{_{\rm GW}}$, the term $(1+t'/\tau'_{_{\rm GW}})^{-1}$ is almost a constant (i.e., $dE_{\rm inj}/dt' \propto {t'}^0$), so that $q=0$ applied in the above analysis is justified. For $B_{\rm s}<10^{15}$ Gauss, the spin-down of the magnetar by a neutrino-driven wind is likely unimportant (Thompson et al. 2004).

\section{Numerical modeling}

Below we present our numerical results to fit the broad-band data.
With the code developed by Fan \& Piran (2006) and Zhang et al. (2006), we take $dE_{\rm inj}/dt'=1.6\times 10^{48}~{\rm erg~s^{-1}}~(1+t'/1000~{\rm s})^{-1}$ for $t'\leq t'_{\rm end}\sim 1000$ s, and $dE_{\rm inj}/dt'=0$ otherwise, to model the afterglow lightcurves (Fig.1). By adopting the following forward shock physical parameters: $\epsilon_{\rm e}=0.15$, $\epsilon_{\rm B}=0.03$, $p=2.3$, $n=0.15~{\rm cm^{-3}}$, $E_{\rm k,0}=2\times 10^{50}$ erg, and $\theta_{\rm j}\approx 0.085$, we show in Fig.1 that the X-ray and optical afterglow lightcurves up to $\sim 1$ day can be reasonably reproduced. {In contrast, the constant energy model (Fig.2 of Fong et al. 2013) over predicts the X-ray flux early on.} Our fit to the radio afterglow is somewhat poor (c.f. Fong et al. 2013), possibly due to the radio scintillation \citep{Goodman1997}.  Here $E_{\rm k,0}$ is the initial kinetic energy of the GRB ejecta and $\theta_{\rm j}$ is the half-opening angle of the GRB ejecta. These shock parameters are consistent with that found in Fong et al. (2013).

The magnetar wind is essentially mildly anisotropic. The wide-beam
wind must run into the merger ejecta. Such an energy injection could
lead the merger ejecta to be a mildly-relativistic speed
\citep{FanXu2006,Zhang2013,Gao2013} and/or to produce a bright ``mergernova'' \citep{Yu2013,Kulkarni2005}.
Adopting the same isotropic-equivalent rate of energy injection
from the magnetar found in modeling GRB afterglow, we fit the
kilonova data using the method delineated in \citet{Yu2013}, where the
dynamical evolution of the merger ejecta is taken into account.
The near infrared data can be well reproduced given an
ejecta mass $M_{\rm ej} \sim 0.02 M_\odot$, an initial velocity of
the ejecta $v_{\rm ej,i}=0.2c$, a magnetar collapsing time $t'_{_{\rm end}}
\sim 1000$ s, and an effective constant opacity
$\kappa_{\rm es}=10\rm cm^2~g^{-1}$. Here the adopted opacity is
much higher than the typical one associated with electron
scattering, because the bound-bound, bound-free, and free-free
transitions of ions could provide more important contributions to
opacity \citep[e.g.][]{Kasen2013,Tanaka2013}.
 The required ejecta mass is at the high-end of simulated ejecta mass
\citep{Hotokezaka2013} for NS-NS mergers. {The kilonova outflow had been accelerated to a velocity $v_{\rm ej,f}=0.36c$ and its total kinetic energy is $\sim 10^{51}$ erg.}

\section{Discussion}
{We have shown that energy injection from a supra-massive magnetar central engine can reproduce the early ($t\lesssim 10^{3}$ s) X-ray and optical afterglow data of the short GRB 130603B. The inferred isotropic-equivalent kinetic energies of both the afterglow and the kilonova are both $\sim 10^{51}$ erg. This is consistent with energy injection from a near-isotropic millisecond magnetar. The relatively small value of the energy budget requires that most energy is carried away via non-electromagnetic signals, and we argue that it is due to gravitational wave radiation with a large deformation of the magnetar.
The proposed Einstein Telescope may be able to detect the required gravitational wave radiation signal, if the source is within a distance $\sim 100$ Mpc \citep{Fan2013}.

The strong evidence of a supra-massive magnetar central engine from
a NS-NS merger event suggests an optimistic prospect of
detecting electromagnetic counterparts of gravitational wave
triggers in the upcoming Advanced LIGO/Virgo era. Since the magnetar
wind is essentially isotropic, a bright early multi-wavelength
afterglow is expected from gravitational wave triggers even without
an associated short GRB \citep{Zhang2013,Gao2013,Yu2013}, which can be
readily detected by wide field X-ray and optical cameras.

Although the supra-massive magnetar model has been widely adopted to interpret the short GRB data, how to produce a short GRB with such a central engine is still a question. One possibility is that the initial nascent neutron star rotates differentially and the magnetic braking and viscosity combine to drive the star to uniform rotation within a time scale $t_{\rm diff}\sim 0.1-1~{\rm s}$, if the surface magnetic field strength of the star reaches $10^{14}-10^{15}$ G \citep{Shapiro2000,Gao2006}. The magnetic activity of the differentially rotating neutron star may be able to drive short but energetic $\gamma-$ray outbursts and thus account for the short GRB prompt emission \citep{Rosswog2007b}. {In the {\it more-speculative} strange quark star scenario, phase transition from neutron matter to strange quark matter may proceed in a short period of time, and the released energy may power a short GRB (Alternatively, the accretion of the disk material onto the strange quark star may power a short GRB).}
In any case, a smoking gun signature of the magneter central engine for short GRBs may be retrieved in the future gravitational wave data.}

\section*{Acknowledgments}
We thank the anonymous referee for insightful suggestions, and R. X. Xu,
B. D. Metzger and W. Fong for helpful communications. This work
was supported in part by 973 Program of China under grant
2013CB837000, National Natural Science of China under grants
11073057, 11273063, and 11103004, and the Foundation for
Distinguished Young Scholars of Jiangsu Province, China (No.
BK2012047). YZF and XFW are also supported by the 100 Talents
programme of Chinese Academy of Sciences. DX acknowledges support
from the ERC-StG grant EGGS-278202 and IDA.

\clearpage

\end{document}